\input{aipcheck}

\documentclass[
    ,final            
  ]
  {aipproc}

\usepackage{axodraw4j}
\usepackage{amssymb}
\usepackage{amsmath}

\layoutstyle{6x9}

\begin{document}

\title{Challenges and Opportunities for the Next Generation of Photon Regeneration Experiments}


\author{Andreas Ringwald}{
  address={Deutsches Elektronen-Synchrotron DESY, Notkestrasse 85, 22607 Hamburg, Germany}
}

\begin{abstract}
Photon regeneration experiments searching for signatures of oscillations of photons into hypothetical very weakly interacting
ultra-light particles, such as axions, axion-like and hidden-sector particles, have improved their sensitivity
considerably in recent years. Important progress in laser and detector technology as well as recycling
of available magnets from accelerators may allow a big further step in sensitivity such that, for
the first time, laser light shining through a wall experiments will explore territory in parameter space
that has not been excluded yet by astrophysics and cosmology. We review these challenges and opportunities
for the next generation experiments.
\end{abstract}

\maketitle

\begin{figure}[ht!]
\scalebox{0.3}[0.3]{
  \begin{picture}(322,141) (95,-63)
    \SetWidth{1.0}
    \SetColor{Black}
    \Photon(96,13)(176,13){7.5}{4}
    \GBox(240,-51)(272,77){0.882}
    \Line[dash,dashsize=2,arrow,arrowpos=0.5,arrowlength=5,arrowwidth=2,arrowinset=0.2](176,13)(336,13)
    \Photon(176,13)(176,-51){7.5}{3}
    \COval(176,-51)(11.314,11.314)(45.0){Black}{White}\Line(170.343,-56.657)(181.657,-45.343)\Line(170.343,-45.343)(181.657,-56.657)

    \Photon(336,13)(336,-51){7.5}{3}
    \Photon(336,13)(416,13){7.5}{4}
    \COval(336,-51)(11.314,11.314)(45.0){Black}{White}\Line(330.343,-56.657)(341.657,-45.343)\Line(330.343,-45.343)(341.657,-56.657)
  \end{picture}
  }
\label{zooalp}
  \hspace*{0.2cm}
\scalebox{0.3}[0.3]{
  \begin{picture}(322,130) (95,-73)
    \SetWidth{1.0}
    \SetColor{Black}
    \Photon(96,2)(176,2){7.5}{4}
    \GBox(240,-62)(272,66){0.882}
    \Photon(336,2)(416,2){7.5}{4}
    \ZigZag(176,2)(336,2){7.5}{8}
    \SetWidth{3.0}
    \Line(160.002,17.998)(191.998,-13.998)\Line(191.998,17.998)(160.002,-13.998)
    \Line(320.002,17.998)(351.998,-13.998)\Line(351.998,17.998)(320.002,-13.998)
  \end{picture}
  }
\label{zoohp}  \hspace*{0.2cm}
\scalebox{0.3}[0.3]{
  \begin{picture}(386,141) (63,-63)
    \SetWidth{1.0}
    \SetColor{Black}
    \GBox(240,-51)(272,77){0.882}
    \Photon(128,13)(64,13){7.5}{3}
    \Arc[arrow,arrowpos=0.5,arrowlength=5,arrowwidth=2,arrowinset=0.2](160,13)(32,270,630)
    \ZigZag(192,13)(320,13){7.5}{6}
    \Arc[arrow,arrowpos=0.5,arrowlength=5,arrowwidth=2,arrowinset=0.2](352,13)(32,270,630)
    \Photon(384,13)(448,13){7.5}{3}
    \Photon(146,-16)(127,-52){7.5}{2}
    \Photon(176,-15)(192,-51){7.5}{2}
    \Photon(336,-15)(320,-51){7.5}{2}
    \Photon(368,-14)(384,-52){7.5}{2}
    \COval(128,-51)(11.314,11.314)(45.0){Black}{White}\Line(122.343,-56.657)(133.657,-45.343)\Line(122.343,-45.343)(133.657,-56.657)
    \COval(192,-51)(11.314,11.314)(45.0){Black}{White}\Line(186.343,-56.657)(197.657,-45.343)\Line(186.343,-45.343)(197.657,-56.657)
    \COval(320,-51)(11.314,11.314)(45.0){Black}{White}\Line(314.343,-56.657)(325.657,-45.343)\Line(314.343,-45.343)(325.657,-56.657)
    \COval(384,-51)(11.314,11.314)(45.0){Black}{White}\Line(378.343,-56.657)(389.657,-45.343)\Line(378.343,-45.343)(389.657,-56.657)
  \end{picture}
  }
\label{zoomcp}
\caption{In LSW experiments, laser photons are sent along a beam onto a wall where they are absorbed.
Some of the photons may converted into WISPs that propagate freely through the wall and reconvert into photons after the wall.
LSW may occur due to various processes beyond the Standard Model:
$\gamma\leftrightarrow$ ALP oscillations in a background magnetic field~\cite{Sikivie:1983ip,Anselm:1987vj,VanBibber:1987rq} (left),
$\gamma\leftrightarrow\gamma^\prime$ oscillations facilitated by a non-zero mass of the hidden photon
($\gamma^\prime$)~\cite{Okun:1982xi} (middle),
and $\gamma\leftrightarrow\gamma^\prime$ oscillations facilitated by virtual mini-charged particles in a background magnetic
field~~\cite{Ahlers:2007rd,Ahlers:2007qf} (right). (From Ref.~\cite{Jaeckel:2010ni}.)
}
\label{Fig:zoo}
\end{figure}
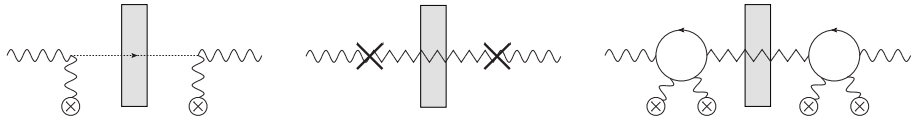
\begin{figure}[ht!]
\includegraphics[width=.33\textwidth]{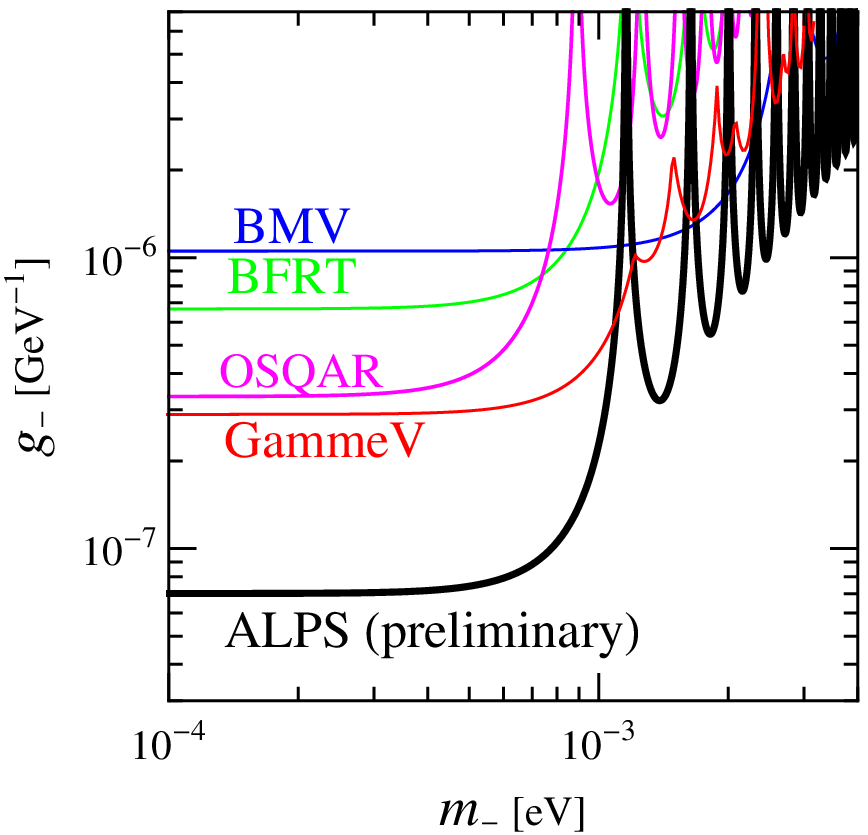}
\hfill
\includegraphics[width=.33\textwidth]{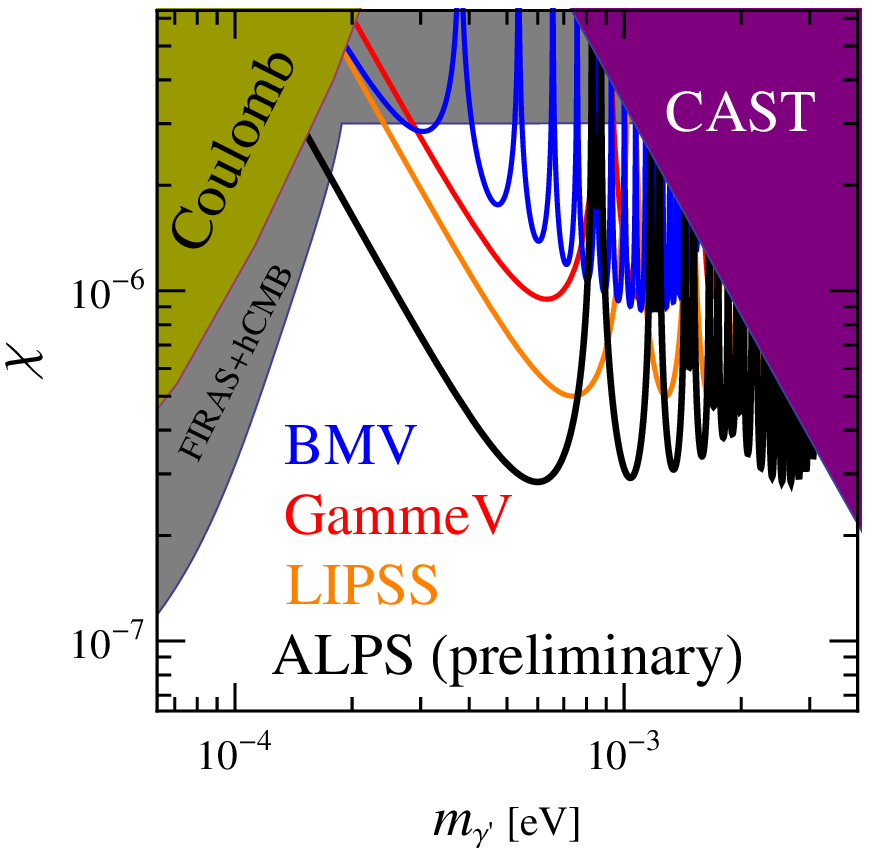}
\hfill
\includegraphics[width=.33\textwidth]{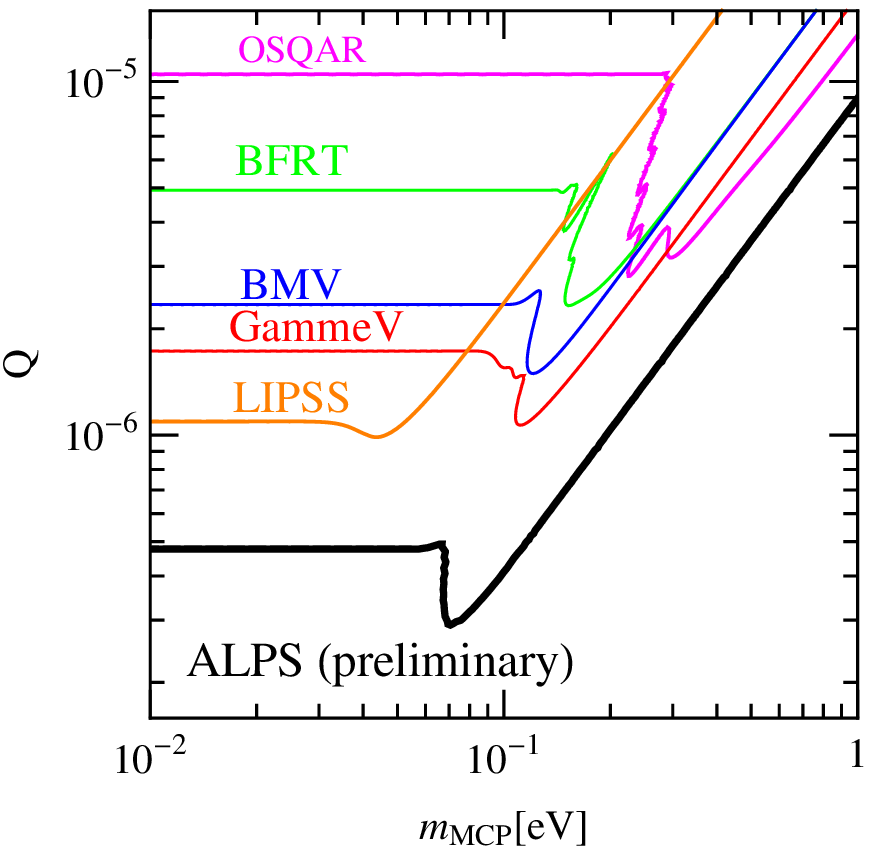}
\caption{Upper bounds from LSW experiments for couplings of
pseudoscalar axion-like particles (two photon coupling $g$; left panel),
massive hidden photons (kinetic mixing $\chi$; middle panel), and massless hidden
photons with an additional minicharged particle (charge $Q=\chi$; right panel).
The results from ALPS are preliminary.
Compilation from Ref.~\cite{ALPS}.}
\label{lswresult}
\end{figure}

Over the last few years, it became more and more clear that precision experiments exploiting low-energy photons
may yield information on particle physics complementary to experiments at high-energy colliders,
in particular on the possible existence of new very weakly interacting sub-eV particles
(WISPs), such as axions~\cite{Weinberg:1977ma}, axion-like particles (ALPs),
and hidden-sector particles (hidden photons~\cite{Okun:1982xi}, minicharged particles~\cite{Holdom:1985ag}),
predicted in many extensions of the Standard Model~\cite{Jaeckel:2010ni,Jaeckel:axions2010}.
The report by the laser polarization experiment PVLAS of the observation of an anomalously large
rotation of the polarization plane of photons after the passage through a magnetic field~\cite{Zavattini:2005tm}
-- which may be interpreted as evidence for photon disappearance due to conversion into
WISPs~\cite{Maiani:1986md,Gies:2006ca,Ahlers:2006iz,Ahlers:2007rd,Ahlers:2007qf} --
provided the impetus for a number of laser light shining through a wall (LSW) experiments. The latter are searching for
photon $\to$ WISP $\to$ photon conversions  (cf. Fig.~\ref{Fig:zoo}) rather than solely for disappearence,
to perform an independent test of the WISP hypothesis~\cite{Robilliard:2007bq,Chou:2007zzc,Pugnat:2007nu,Afanasev:2008jt,Fouche:2008jk,Afanasev:2008fv,Ehret:2009sq}
und to improve the constraints from the pioneering experiment BFRT~\cite{Cameron:1993mr} by about an order of magnitude in
the WISP--photon coupling (cf. Fig.~\ref{lswresult}).
Moreover, the momentum gained by these experiments towards the establishment of a new
low-energy frontier of particle physics turned out to be conserved even though the original motivation disappeared: the PVLAS collaboration could not reestablish their first observation after an upgrade of their apparatus~\cite{Zavattini:2007ee}. This is in-line with the finding of the above
mentioned LSW experiments.

Now, the planning for the next generation of photon regeneration experiments has started.
At this stage, it seems to be very helpful to identify targets in WISP parameter space upon which
the next generation of experiments can shoot. In this context, one can clearly identify both
\begin{itemize}
\item {\bf challenges:} increase sensitivity beyond astro, cosmo, and other lab bounds, and
\item {\bf opportunities:} test WISP interpretation of hints for cosmic photon regeneration,
\end{itemize}
that we will discuss in detail in the following.

\begin{figure}[t!]
\centerline{
\includegraphics[angle=-90,width=.75\textwidth]{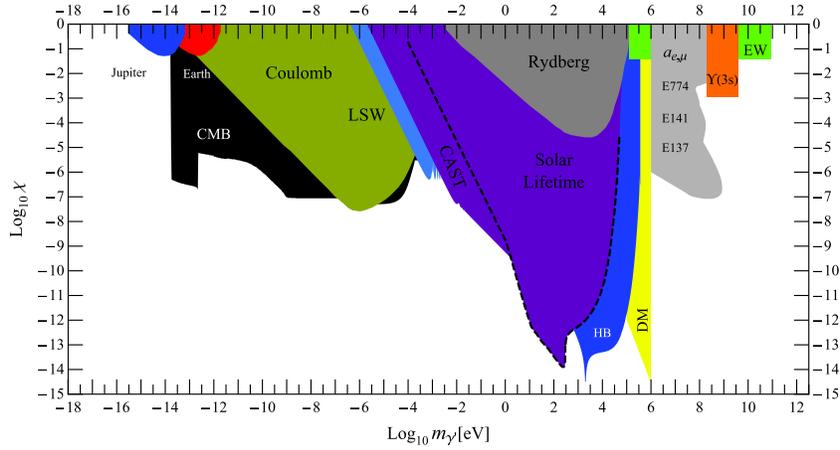}
}
\caption{
Summary of astrophysical, cosmological and laboratory constraints for hidden photons (kinetic mixing
$\chi$ vs. mass $m_{\gamma^\prime}$) (Adapted from Ref.~\cite{Goodsell:2009xc}, where also details can be found.).
}
\label{Fig:hp_astro}
\end{figure}

For hidden photons, laser LSW experiments are in a comfortable position, as is illustrated in
Figs.~\ref{lswresult} (middle) and \ref{Fig:hp_astro}: already by now, they are exploring previously
untouched parameter space, bearing therefore the greatest immediate discovery potential.
The cosmo constraint arising from the upper limit on the effective number of relativistic degrees
of freedom contributing to the cosmic radiation density in the era between big bang
nucleosynthesis and recombination~\cite{Jaeckel:2008fi} (grey area in Fig.~\ref{lswresult} (middle)) as well as the constraint arising from a search for photon regeneration
due to solar hidden photons in the CAST helioscope~\cite{Redondo:2008aa} (purple area in Fig.~\ref{lswresult} (middle)) are not competitive with LSW limits in the $\sim$ meV
mass range.

This is, however, only true if there is no light physical hidden Higgs particle involved, i.e.
if the hidden photon gets its mass from a St\"uckelberg mechanism.
Otherwise, if the hidden photon mass arises via a Higgs mechanism,
the physical hidden Higgs effectively acts as a minicharged particle, with charge $Q= \chi e_h/e$,
where $e_h$ is the gauge coupling of the hidden photon, and the
strong astro bound $Q\lesssim 10^{-14}$, for a sub-keV hidden Higgs mass,
inferred from the lifetime of red giants applies~\cite{Davidson:2000hf}.
\begin{figure}[b!]
\centerline{
\includegraphics[angle=-90,width=.75\textwidth]{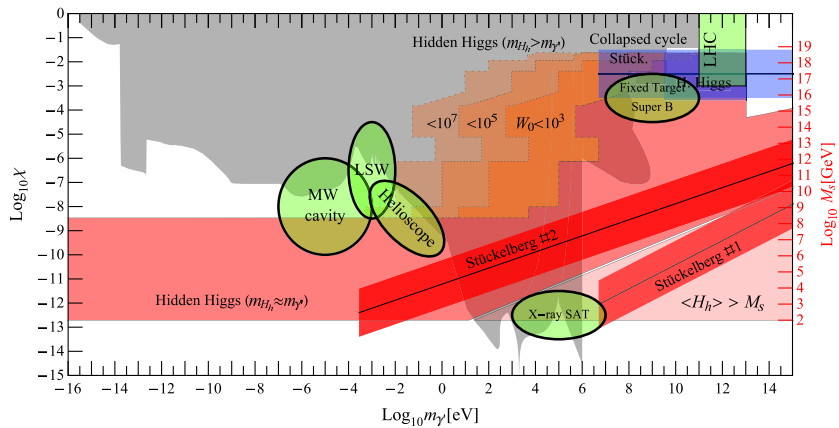}
}
\caption{
Prediction of hidden photon kinetic mixing $\chi$ with the visible photon vs. its mass $m_{\gamma^\prime}$
from LARGE volume string compactifications. The grey area is excluded by hidden photon searches alone.
The bright red region predicted for hyperweak hidden photons whose mass arises from a hidden Higgs mechanism
takes already into account the astro and cosmo constraints from minicharged particles~\cite{Davidson:2000hf}.
(Compilation from Ref.~\cite{Goodsell:2009xc}.).}
\label{Fig:hp_string}
\end{figure}
In particularly well motivated LARGE volume string compactifications, the gauge coupling of the hidden photon can be hyperweak,
i.e. diluted due to the volume of the
extra dimensions, to $e_h\sim 10^{-6}$, for a volume corresponding to an intermediate string scale
$M_s\sim 10^{9}$~GeV~\cite{Burgess:2008ri}.
Therefore, the limit on minicharged particles excludes $\chi\gtrsim {\rm few}\times 10^{-9}$, at low masses,
$m_{\gamma^\prime}\sim m_{H_h}\lesssim$~keV (cf. Fig.~\ref{Fig:hp_string}). Thus, the discovery potential
for hidden photons would be increased dramatically if we were able to probe such low values of $\chi$ with the
next generation of laser LSW experiments.

Fortunately, this seems doable.
The current state-of-the-art LSW experiment
ALPS, exploiting an optical resonator at the generation side of the experiment, resulting in a
power of $\sim 1.2$~kW available for $\gamma\to$~WISP conversions, established an upper limit
$P_{\rm LSW}\lesssim {\rm few}\times 10^{-25}$ on the LSW probability, corresponding to an upper
limit $\chi\lesssim {\rm few}\times 10^{-7}$ in the meV mass range. Exploiting additionally a high finesse ($\sim 10^4$)
optical resonator also on the regeneration side of the experiment~\cite{Hoogeveen:1990vq,Sikivie:2007qm} and a single-photon
counter, together with an increased power buildup, by a factor of $\sim 100$, on the generation side,
it seems possible to improve the sensitivity on the LSW probability by $\sim 4+2+2 = 8$ orders of magnitude,
corresponding to an improved sensitivity in $\chi$ by $\sim 8/4 = 2$ orders of magnitude, down to the most interesting
values, $\chi\sim {\rm few}\times 10^{-9}$.
Such values, at somewhat smaller masses, can also be probed by microwave cavity variants of the LSW
technique~\cite{Hoogeveen:1992uk,Jaeckel:2007ch,Caspers:2009cj}, which are currently set
up~\cite{Povey:2010hs,Baker:axions2010,Rybka:axions2010}, and,
at somewhat larger masses, by especially designed helioscopes to search for solar hidden photons~\cite{Gninenko:2008pz},
which are also under consideration (see, e.g., Ref.~\cite{Minowa:axions2010}).

Let us turn now to axions and ALPs. Although much less model dependent~\cite{Jaeckel:2006xm},
the values of the two photon coupling $g$ of ALPs probed by the
current generation of LSW experiments, $g\gtrsim {\rm few}\times 10^{-8}$~GeV$^{-1}$, for masses below
\begin{figure}[h!]
\centerline{
\includegraphics[width=.5\textwidth]{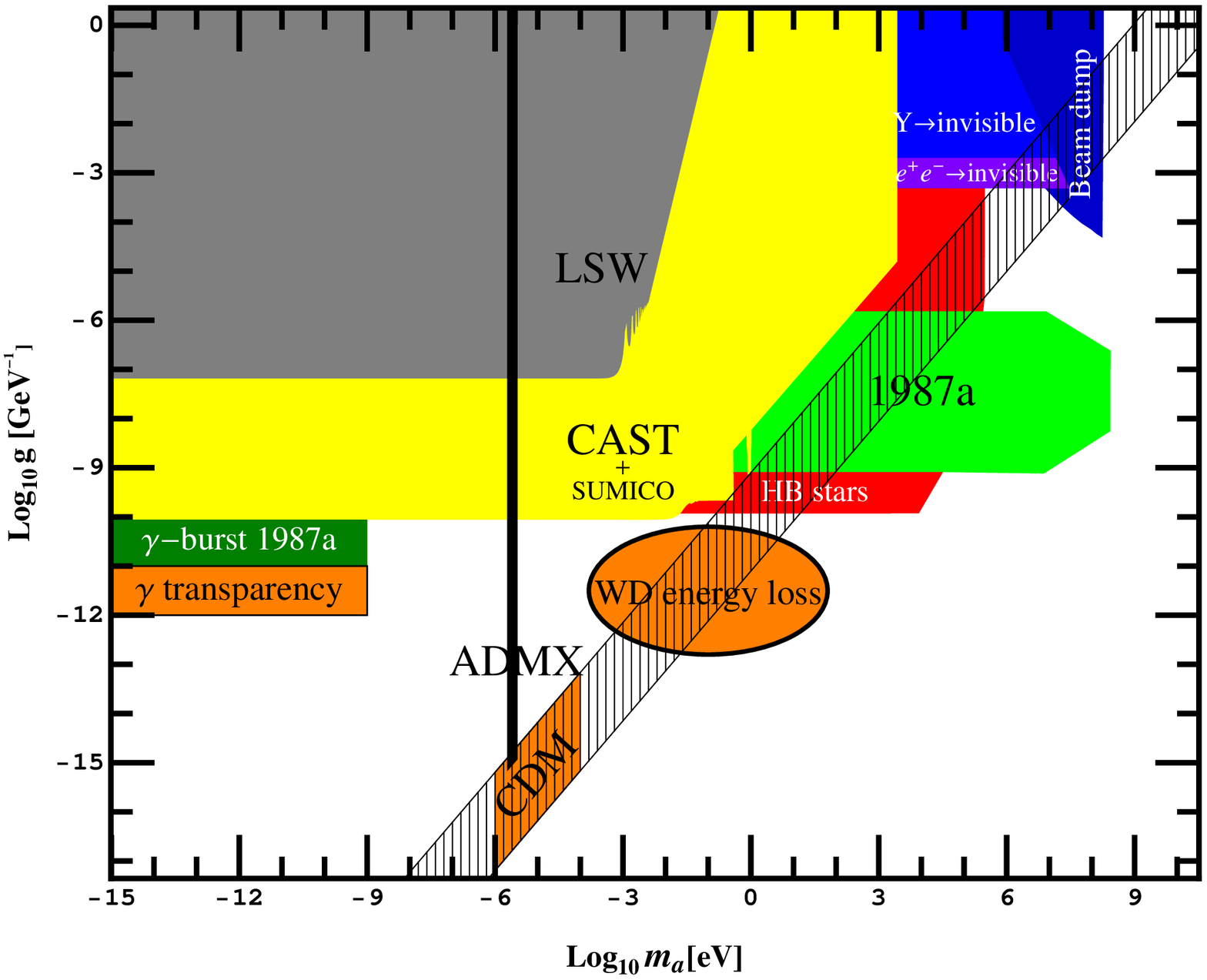}
\includegraphics[width=.5\textwidth]{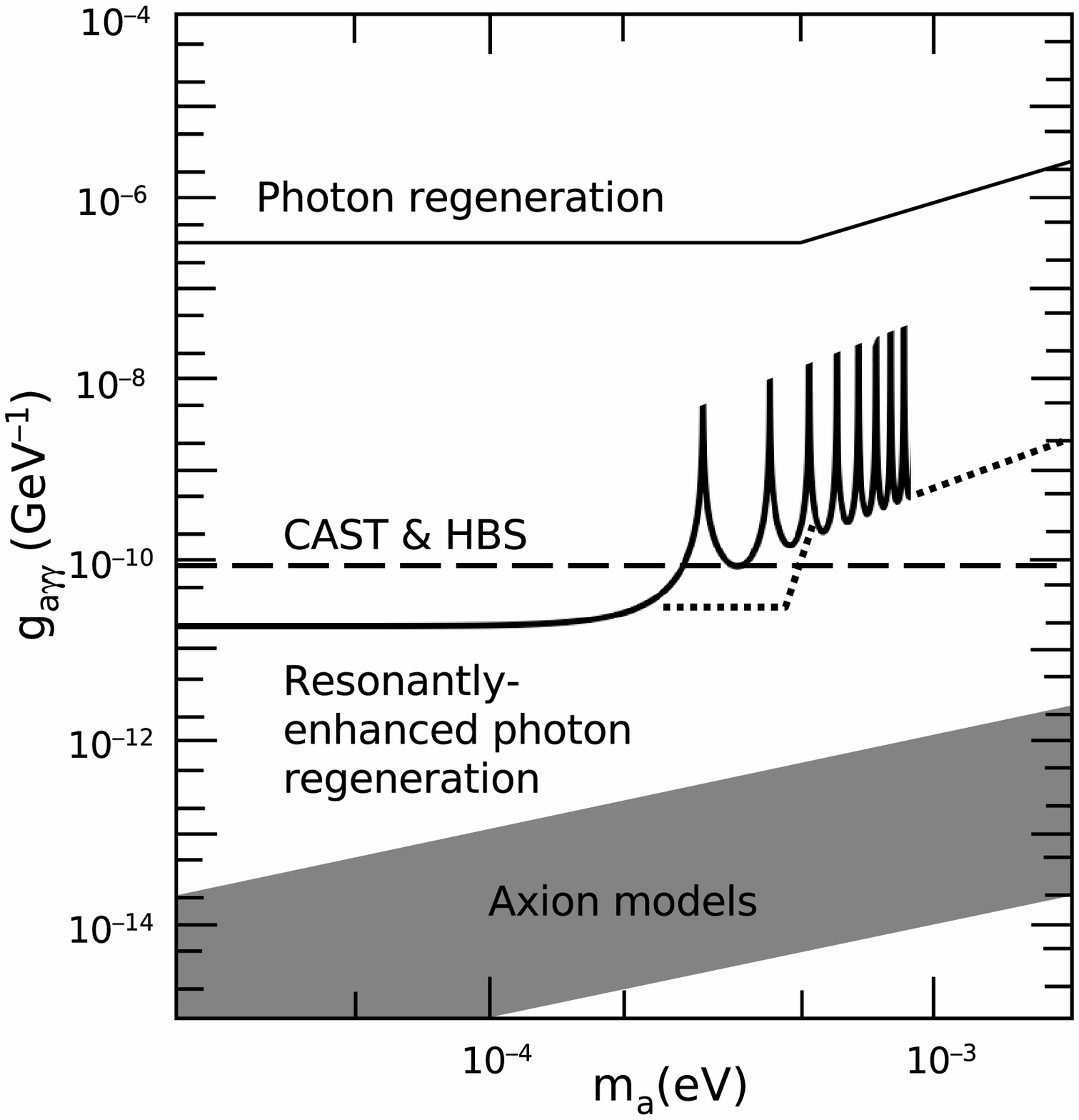}
}
\caption{{\em Left:} Summary of cosmological and astrophysical constraints
for axion-like-particles (two
photon coupling $g$ vs. mass $m_{a}$ of the
ALP).
Note that the mass region, where the axion can
be the cold dark matter (the orange regions labeled  ``CDM"),
can be extended towards smaller masses by anthropic reasoning.
Also other areas with interesting astrophysical hints, e.g. the one for
a non-standard energy loss in white dwarfs~\cite{Isern:2008nt} or the one 
for an anomalous $\gamma$-ray transparency of the universe~\cite{De Angelis:2007dy,DeAngelis:2008sk}, 
are marked in orange.
The parameter range for the axion is shown hatched.
Note
that the limit from the microwave cavity axion dark matter search experiment
ADMX~\cite{Duffy:2006aa} is valid only under the assumption that the
local density of ALPs at earth is given by the dark matter density.
(Compilation from Ref.~\cite{Jaeckel:2010ni}, where also details can be found.)
{\em Right:} Prospected sensitivity of a laser LSW experiment exploiting
6+6 Tevatron magnets and resonantly enhanced photon regeneration~\cite{Mueller:2009wt}.
}\label{Fig:axions_astro}
\end{figure}
an meV, falls short, by nearly three orders of magnitude,
to the strong limits established by lifetime considerations of
horizontal branch stars~\cite{Raffelt:1985nk,Raffelt:1987yu} and by limits on photon regeneration due to
solar ALPs reported by the helioscopes CAST~\cite{Andriamonje:2007ew} and SUMICO~\cite{Inoue:2008zp}
(cf. Fig.~\ref{Fig:axions_astro} (left)). Here, the next generation of LSW experiments has to gain about
three orders of magnitude in the coupling to start to enter in previously
unexplored territory. In addition to above mentioned improvements from the laser and detector
side, one has to increase $B\times L$, the magnitude times the length of the magnetic field region,
by one order of magnitude compared to the current experiments,
e.g. by exploiting 5+5 HERA magnets at ALPS, instead of the current 1/2+1/2 configuration.
With such improvements, a sensitivity in the $g\sim {\rm few}\times 10^{-11}$~GeV$^{-1}$ range, for
light ALPs, $m_\phi\ll$~meV, should be
achievable~\cite{Mueller:2009wt,Lindner:axions2010,Meier:axions2010,Mueller:axions2010}.
For the sensitivity of a similar setup proposed in Ref.~\cite{Mueller:2009wt}, exploiting 6+6 Tevatron magnets,
see Fig.~\ref{Fig:axions_astro} (right).

An even wider range of opportunities for discovery would open up if the sensitivity in $g$
can be improved even more, by one order of magnitude, down to $g\sim {\rm few}\times 10^{-12}$~GeV$^{-1}$,
possibly by a combination of laser and magnet upgrades.

First of all, ALPs with such a coupling may
be motivated from a top-down perspective arising from string theory. In fact, massless ALPs, with
coupling to photons in the $g\sim \alpha/M_s\sim 10^{-12}\div 10^{-11}$~GeV$^{-1}$ range, could occur
naturally in string compactifications with an intermediate string scale
$M_s\sim 10^9\div 10^{10}$~GeV.

Secondly, there are a number of puzzling astronomical observations which
may be commonly explained by cosmic photon oscillations into  very light ALPs with $g$ in the above range
(cf. also Fig.~\ref{Fig:axions_astro} (left)). Indeed, photons emitted by distant sources and propagating
through cosmic magnetic fields can oscillate into ALPs, with a number of consequences in different
situations (see, e.g., Refs.~\cite{Csaki:2001yk,Csaki:2003ef,Mirizzi:2007hr,Hooper:2007bq,Hochmuth:2007hk,Payez:2008pm,Fairbairn:2009zi,%
Mirizzi:2009nq,Chou:axions2010,Payez:axions2010,Redondo:axions2010}).
Interestingly, ALPs may leave their imprints in luminosity relations of active galactic
nuclei~\cite{Burrage:2009mj,Burrage:axions2010}.
In fact, mixing between photons and ALPs in the random magnetic fields in galaxy clusters
induces a characteristic scatter in the relations of X-ray vs. optical luminosities of compact sources
in these clusters. Evidence for such an effect has recently been found in an analysis of
luminosity relations of about two hundred active galactic nuclei, providing
a strong hint for the possible existence of a very light, $m_\phi\lesssim 10^{-12}$~eV,
ALP, with a coupling in the $g\sim 10^{-12}\div 10^{-11}$~GeV$^{-1}$ range.

This range is also the sensitivity of another astrophysical probe of ALPs, namely
the spectra of cosmologically distant TeV $\gamma$-ray sources.
In fact, recent observations of a few of them by ground-based gamma ray telescopes
have revealed a surprising degree of transparency of the universe to very high-energy photons, which seems
to point to less absorption due to pair production, may be due to a less dense extragalactic background light and/or
to a harder injection spectrum at the sources than initially thought. However, there is also the intriguing
possibility to explain this puzzle through photon $\leftrightarrow$ ALPs oscillations in the
cosmic magnetic fields, again requiring a coupling in the $g\sim 10^{-12}\div 10^{-11}$~GeV$^{-1}$
range~\cite{De Angelis:2007dy,DeAngelis:2008sk} (cf. Fig.~\ref{Fig:axions_astro} (left)).
The present status of this affair is far from conclusive, however. It seems that
much more data from many more quite distant TeV gamma sources along different directions in the sky has to be
collected before one may be able to perform a systematic search for hints of ALPs~\cite{Mirizzi:2009aj}.
For this increase in statistics, we have to wait, however, for the realization of the big
TeV gamma ray array CTA. It would be great, if we were able to probe the same range of parameters
even earlier in the laboratory, by laser light shining through a wall!


\bibliographystyle{aipproc}   

\begin{thebibliography}{9}

\bibitem{Weinberg:1977ma}
  S.~Weinberg,
  Phys.\ Rev.\ Lett.\  {\bf 40}, 223 (1978);
%
  F.~Wilczek,
  Phys.\ Rev.\ Lett.\  {\bf 40}, 279 (1978).

\bibitem{Okun:1982xi}
  L.~B.~Okun,
  Sov.\ Phys.\ JETP {\bf 56}, 502 (1982)
  [Zh.\ Eksp.\ Teor.\ Fiz.\  {\bf 83}, 892 (1982)].

\bibitem{Holdom:1985ag}
  B.~Holdom,
  Phys.\ Lett.\  B {\bf 166}, 196 (1986).

\bibitem{Jaeckel:2010ni}
  J.~Jaeckel and A.~Ringwald,
  arXiv:1002.0329 [hep-ph].

\bibitem{Jaeckel:axions2010}
J.~Jaeckel, these proceedings

\bibitem{Zavattini:2005tm}
  E.~Zavattini {\it et al.}  [PVLAS Collaboration],
  Phys.\ Rev.\ Lett.\  {\bf 96}, 110406 (2006).

\bibitem{Maiani:1986md}
  L.~Maiani, R.~Petronzio and E.~Zavattini,
  Phys.\ Lett.\  B {\bf 175}, 359 (1986).

\bibitem{Gies:2006ca}
  H.~Gies, J.~Jaeckel and A.~Ringwald,
  Phys.\ Rev.\ Lett.\  {\bf 97}, 140402 (2006).

\bibitem{Ahlers:2006iz}
  M.~Ahlers, H.~Gies, J.~Jaeckel and A.~Ringwald,
  Phys.\ Rev.\  D {\bf 75}, 035011 (2007).

\bibitem{Ahlers:2007rd}
  M.~Ahlers, H.~Gies, J.~Jaeckel, J.~Redondo and A.~Ringwald,
  Phys.\ Rev.\  D {\bf 76}, 115005 (2007).

\bibitem{Ahlers:2007qf}
  M.~Ahlers, H.~Gies, J.~Jaeckel, J.~Redondo and A.~Ringwald,
  Phys.\ Rev.\  D {\bf 77}, 095001 (2008).

\bibitem{Sikivie:1983ip}
  P.~Sikivie,
  Phys.\ Rev.\ Lett.\  {\bf 51}, 1415 (1983)
  [Erratum-ibid.\  {\bf 52}, 695 (1984)].

\bibitem{Anselm:1987vj}
  A.~A.~Anselm,
  Phys.\ Rev.\  D {\bf 37}, 2001 (1988).

\bibitem{VanBibber:1987rq}
  K.~Van Bibber {\it et al.}, 
  Phys.\ Rev.\ Lett.\  {\bf 59}, 759 (1987).

\bibitem{Robilliard:2007bq}
  C.~Robilliard {\it et al.}, 
  Phys.\ Rev.\ Lett.\  {\bf 99}, 190403 (2007).

\bibitem{Chou:2007zzc}
  A.~S.~Chou {\it et al.}  [GammeV (T-969) Collaboration],
  Phys.\ Rev.\ Lett.\  {\bf 100}, 080402 (2008).

\bibitem{Pugnat:2007nu}
  P.~Pugnat {\it et al.}  [OSQAR Collaboration],
  Phys.\ Rev.\  D {\bf 78}, 092003 (2008).

\bibitem{Afanasev:2008jt}
  A.~Afanasev {\it et al.},
  Phys.\ Rev.\ Lett.\  {\bf 101}, 120401 (2008).

\bibitem{Fouche:2008jk}
  M.~Fouche {\it et al.},
  Phys.\ Rev.\  D {\bf 78}, 032013 (2008).

\bibitem{Afanasev:2008fv}
  A.~Afanasev {\it et al.},
  Phys.\ Lett.\  B {\bf 679}, 317 (2009).

\bibitem{Ehret:2009sq}
  K.~Ehret {\it et al.}  [ALPS collaboration],
  Nucl.\ Instrum.\ Meth.\  A {\bf 612}, 83 (2009).

\bibitem{Cameron:1993mr}
  R.~Cameron {\it et al.},
  Phys.\ Rev.\  D {\bf 47}, 3707 (1993).

\bibitem{ALPS}
  A. Lindner, J. Redondo, for the [ALPS Collaboration], priv. comm.
  and publ. (DESY 10-030) in prep.

\bibitem{Zavattini:2007ee}
  E.~Zavattini {\it et al.}  [PVLAS Collaboration],
  Phys.\ Rev.\  D {\bf 77}, 032006 (2008).

\bibitem{Goodsell:2009xc}
  M.~Goodsell, J.~Jaeckel, J.~Redondo and A.~Ringwald,
  JHEP {\bf 0911}, 027 (2009).

\bibitem{Jaeckel:2008fi}
  J.~Jaeckel, J.~Redondo and A.~Ringwald,
  Phys.\ Rev.\ Lett.\  {\bf 101}, 131801 (2008).

\bibitem{Redondo:2008aa}
  J.~Redondo,
  JCAP {\bf 0807}, 008 (2008).

\bibitem{Davidson:2000hf}
  S.~Davidson, S.~Hannestad and G.~Raffelt,
  JHEP {\bf 0005}, 003 (2000).

\bibitem{Burgess:2008ri}
  C.~P.~Burgess {\it et al.}, 
  JHEP {\bf 0807}, 073 (2008).

\bibitem{Hoogeveen:1990vq}
  F.~Hoogeveen and T.~Ziegenhagen,
  Nucl.\ Phys.\  B {\bf 358}, 3 (1991).

\bibitem{Sikivie:2007qm}
  P.~Sikivie, D.~B.~Tanner and K.~van Bibber,
  Phys.\ Rev.\ Lett.\  {\bf 98}, 172002 (2007).

\bibitem{Hoogeveen:1992uk}
  F.~Hoogeveen,
  Phys.\ Lett.\  B {\bf 288}, 195 (1992).

\bibitem{Jaeckel:2007ch}
  J.~Jaeckel and A.~Ringwald,
  Phys.\ Lett.\  B {\bf 659}, 509 (2008).

\bibitem{Caspers:2009cj}
  F.~Caspers, J.~Jaeckel and A.~Ringwald,
  JINST {\bf 4}, P11013 (2009).

\bibitem{Povey:2010hs}
  R.~Povey, J.~Hartnett and M.~Tobar,
  arXiv:1003.0964 [hep-ex].

\bibitem{Baker:axions2010}
O.~K.~Baker, these proceedings

\bibitem{Rybka:axions2010}
G.~Rybka, these proceedings

\bibitem{Gninenko:2008pz}
  S.~N.~Gninenko and J.~Redondo,
  Phys.\ Lett.\  B {\bf 664}, 180 (2008).

\bibitem{Minowa:axions2010}
M.~Minowa, these proceedings

\bibitem{Jaeckel:2006xm}
  J.~Jaeckel, E.~Masso, J.~Redondo, A.~Ringwald and F.~Takahashi,
  Phys.\ Rev.\  D {\bf 75}, 013004 (2007).

\bibitem{Raffelt:1985nk}
  G.~G.~Raffelt,
  Phys.\ Rev.\  D {\bf 33}, 897 (1986).

\bibitem{Raffelt:1987yu}
  G.~G.~Raffelt and D.~S.~P.~Dearborn,
  Phys.\ Rev.\  D {\bf 36}, 2211 (1987).

\bibitem{Andriamonje:2007ew}
  S.~Andriamonje {\it et al.}  [CAST Collaboration],
  JCAP {\bf 0704}, 010 (2007).

\bibitem{Inoue:2008zp}
  Y.~Inoue {\it et al.}, 
  Phys.\ Lett.\  B {\bf 668}, 93 (2008).

\bibitem{Isern:2008nt}
  J.~Isern, E.~Garcia-Berro, S.~Torres and S.~Catalan,
  arXiv:0806.2807 [astro-ph].

\bibitem{De Angelis:2007dy}
  A.~De Angelis, O.~Mansutti and M.~Roncadelli,
  Phys.\ Rev.\  D {\bf 76}, 121301 (2007).

\bibitem{DeAngelis:2008sk}
  A.~De Angelis, O.~Mansutti, M.~Persic and M.~Roncadelli,
  arXiv:0807.4246 [astro-ph].

\bibitem{Duffy:2006aa}
  L.~D.~Duffy {\it et al.},
  Phys.\ Rev.\  D {\bf 74}, 012006 (2006).

\bibitem{Mueller:2009wt}
  G.~Mueller, P.~Sikivie, D.~B.~Tanner and K.~van Bibber,
  Phys.\ Rev.\  D {\bf 80}, 072004 (2009).

\bibitem{Lindner:axions2010}
A.~Lindner, these proceedings

\bibitem{Meier:axions2010}
T.~Meier, these proceedings

\bibitem{Mueller:axions2010}
G.~Mueller, these proceedings

\bibitem{Csaki:2001yk}
  C.~Csaki, N.~Kaloper and J.~Terning,
  Phys.\ Rev.\ Lett.\  {\bf 88}, 161302 (2002).

\bibitem{Csaki:2003ef}
  C.~Csaki, N.~Kaloper, M.~Peloso and J.~Terning,
  JCAP {\bf 0305}, 005 (2003).

\bibitem{Mirizzi:2007hr}
  A.~Mirizzi, G.~G.~Raffelt and P.~D.~Serpico,
  Phys.\ Rev.\  D {\bf 76}, 023001 (2007).

\bibitem{Hooper:2007bq}
  D.~Hooper and P.~D.~Serpico,
  Phys.\ Rev.\ Lett.\  {\bf 99}, 231102 (2007).

\bibitem{Hochmuth:2007hk}
  K.~A.~Hochmuth and G.~Sigl,
  Phys.\ Rev.\  D {\bf 76}, 123011 (2007).

\bibitem{Payez:2008pm}
  A.~Payez, J.~R.~Cudell and D.~Hutsemekers,
  AIP Conf.\ Proc.\  {\bf 1038}, 211 (2008).

\bibitem{Fairbairn:2009zi}
  M.~Fairbairn, T.~Rashba and S.~V.~Troitsky,
  arXiv:0901.4085 [astro-ph.HE].

\bibitem{Mirizzi:2009nq}
  A.~Mirizzi, J.~Redondo and G.~Sigl,
  JCAP {\bf 0908}, 001 (2009).

\bibitem{Chou:axions2010}
A.~Chou, these proceedings

\bibitem{Payez:axions2010}
A.~Payez, these proceedings

\bibitem{Redondo:axions2010}
J.~Redondo, these proceedings

\bibitem{Burrage:2009mj}
  C.~Burrage, A.~C.~Davis and D.~J.~Shaw,
  Phys.\ Rev.\ Lett.\  {\bf 102}, 201101 (2009).

\bibitem{Burrage:axions2010}
C.~Burrage, these proceedings

\bibitem{Mirizzi:2009aj}
  A.~Mirizzi and D.~Montanino,
  JCAP {\bf 0912}, 004 (2009).


\end{thebibliography}


\end{document}